\documentclass[onecolumn,amsmath,amssymb,11pt,superscriptaddress,nofootinbib]{revtex4}
\pdfoutput=1

\usepackage[latin1]{inputenc}
\usepackage[english]{babel}
\usepackage{amssymb}
\usepackage{amsmath}
\usepackage{amsthm}
\usepackage[]{graphicx}
\usepackage[]{subfigure}
\usepackage{tensor}
\usepackage{color}
\usepackage{cancel}
\usepackage{setspace}
\usepackage{fancyhdr}
\usepackage[bookmarks,linktocpage, colorlinks=true, plainpages = false, citecolor = blue,  linkcolor=blue, urlcolor = blue, filecolor = blue]{hyperref} 

\begin{document}

\allowdisplaybreaks
\begin{titlepage}

\title{The Wave Function of Simple Universes, Analytically Continued From Negative to Positive Potentials \vspace{.3in}}

\author{Jean-Luc Lehners\\ {\it Max--Planck--Institute for Gravitational Physics (Albert--Einstein--Institute), 14476 Potsdam, Germany \\ jlehners@aei.mpg.de}}


\begin{abstract}
\vspace{.3in} \noindent 
We elaborate on the correspondence between the canonical partition function in asymptotically AdS universes and the no-boundary proposal for positive vacuum energy. For the case of a pure cosmological constant, the analytic continuation of the AdS partition function is seen to define the no-boundary wave function (in dS) uniquely in the simplest minisuperspace model. A consideration of the AdS gravitational path integral implies that on the dS side, saddle points with Hawking-Moss/Coleman-De Luccia-type tunnelling geometries are irrelevant. This implies that simple topology changing geometries do not contribute to the nucleation of the universe. The analytic AdS/dS equivalence holds up once tensor fluctuations are added. It also works, at the level of the saddle point approximation, when a scalar field with a mass term is included, though in the latter case, it is the mass that must be analytically continued. Our results illustrate the emergence of time from space by means of a Stokes phenomenon, in the case of positive vacuum energy. Furthermore, we arrive at a new characterisation of the no-boundary condition, namely that there should be no momentum flux at the nucleation of the universe.
\end{abstract}
\maketitle

\end{titlepage}

\tableofcontents

\section{Introduction} \label{sec:introduction}

The close connections between gravitational path integrals with Anti-de Sitter (AdS) and with de Sitter (dS) asymptotics have been studied for some time, see e.g. \cite{Witten:2001kn,Strominger:2001pn,Maldacena:2002vr,McFadden:2009fg,Hertog:2011ky}. Usually these connections have been explored at the level of saddle point geometries, i.e. at the level of solutions of the classical equations of motion. Progress was made recently in \cite{DiTucci:2020weq}, where the correspondence could be extended very explicitly to the full path integral, albeit with metrics restricted by symmetry assumptions (i.e. in minisuperspace). Crucial for the successful implementation of the path integral was the imposition of a regularity condition in the bulk, required such that the asymptotically AdS geometries cap off smoothly. This regularity condition took the form of a condition on the momentum of the scale factor of the universe, i.e. the path integral needed to be defined with a Neumann condition on one end, and a Dirichlet condition at the ``outer'' boundary, where the metric was held fixed. It is important that this choice of boundary conditions was not arbitrary, but was required to match established thermodynamic properties of black holes in AdS, such as the Hawking-Page phase transition \cite{Hawking:1982dh}. Interestingly, the required momentum condition turns out to be identical to that used in recent path integral implementations of the no-boundary proposal\footnote{The original descriptions of the no-boundary proposal can be found in \cite{Hawking:1981gb,Hartle:1983ai,Hawking:1983hj}.} \cite{DiTucci:2019dji,DiTucci:2019bui} (in those works, the momentum condition was imposed in order to eliminate saddle points with unstable perturbations), thus demonstrating a direct correspondence between path integrals with AdS asymptotics and the no-boundary proposal in dS. 

The aim of the present paper is two-fold: first to take a closer look at the implications of the above-described correspondence for simple cosmologies, and second to extend the correspondence to tensor perturbations and scalar fields. 

Regarding the first point, one can use the analytic properties of the exact AdS wave function in order to define the no-boundary wave function uniquely (at least in the simple setting studied here). Moreover, the analytic continuation indicates that there is no contribution to the no-boundary wave function from simple geometries with a different topology. In particular, there is no contribution to the nucleation of the universe from Hawking-Moss or Coleman-De Luccia instantons \cite{Hawking:1981fz,Coleman:1980aw}. Furthermore, time is seen to arise from space via a Stokes phenomenon, where the path integral is initially dominated by a single saddle point which, after joining with an irrelevant saddle, splits into two complex relevant saddles. This is the basic difference between AdS path integrals, where no Stokes phenomenon occurs, and dS path integrals, where the Stokes phenomenon allows asymptotically Lorentzian geometries to become relevant, and thus allows one space direction to morph into a time direction.

The extension to perturbations and scalar fields can only be described at the level of the saddle point approximation, but there it is seen to hold. We recover the well known result that for tensor perturbations the Bunch-Davies ground state is implied. We illustrate, however, the rather striking differences with standard inflationary calculations. In particular, on the saddle points (as represented in the minisuperspace path integral, i.e. with a constant, complex, lapse function) the perturbations do not oscillate when the universe is smaller than the dS radius, and they are massively suppressed at early ``times'' when the wave number is high. This demonstrates  how the no-boundary proposal manages to significantly smooth out fluctuations. When the universe is small, cosmological fluctuations behave such as those in AdS, and not such as those in the flat slicing of de Sitter space, as commonly described for classical inflationary models. This may have implications for considerations of trans-Planckian issues. The AdS/dS equivalence continues to work when deformed to include a scalar field with a mass term, though in this case one must analytically continue the mass. Here also the dominant geometries remain Euclidean for negative potentials, and become complex (asymptotically Lorentzian) for positive potentials. Thus a Stokes phenomenon is again responsible for the emergence of time from space.

A note on conventions: we set $8\pi G=1,$ and use Lorentzian metrics with mostly plus signature. In intermediate calculations, $\hbar$ is set to unity, but restored at the end. We work in $4$ dimensions.

\section{Pure cosmological constant} \label{sec:cc}

The theory we consider at first is that of gravity with a cosmological constant $\Lambda,$ with action
\begin{align}
S=\frac{1}{2}\int d^4 x \sqrt{-g} \left( R - 2\Lambda \right) + \int d^3y \sqrt{g}K\mid_{outer \, boundary}\,, \label{act}
\end{align}
where $K$ is the trace of the extrinsic curvature on the outer boundary; this term will be discussed in more detail below. We will be interested in both positive and negative values of $\Lambda$ -- in fact, we will consider an analytic continuation in $\Lambda,$ so that it will be useful to think of $\Lambda$ as a complex parameter in general.
With the metric \cite{Halliwell:1988ik}
\begin{align}
ds^2 = - \frac{N^2}{q}dt^2 + q \, d\Omega_3^2\,, \label{metric}
\end{align}
where $N$ is the lapse function, $q(t)$ the square of the scale factor and $d\Omega_3^2$ the metric on a unit 3-sphere of volume $2\pi^2,$ this simple minisuperspace model is described by the Lagrangian
\begin{align}
L = 2\pi^2\left[ - \frac{3}{4N}\dot{q}^2 + 3N - N \Lambda q\right]\,, \label{Lag}
\end{align}
where a dot denotes a derivative w.r.t. $t.$ The canonical momentum $p$ corresponding to $q$ is given by
\begin{align}
p = \frac{\partial L}{\partial \dot{q}} = -   \frac{3 \pi^2}{N}  \dot{q}\,.
\end{align}
Consequently the Hamiltonian takes the simple form
\begin{align}
H =& \dot{q} p -L= -\frac{N}{6 \pi^2} \left[ p^2 + 12 \pi^4 (3 -\Lambda q)\right]=N\hat{H}\,. \label{Hamiltonian}
\end{align}
The Wheeler-DeWitt equation then corresponds to the operator version of the classical constraint,
\begin{align}
\hat{H} \Psi = 0\,,
\end{align}
where $\Psi$ is the wave function of the universe.

The canonical commutation relation $[q,p]=i$ can be implemented in different ways. In the position/field representation, we may represent the momentum operator by the substitution $p \mapsto \hat{p} = - i  \frac{\partial}{\partial q}$, leading to the Wheeler-DeWitt equation 
\begin{align}
\hat{H}_{(q)} \Psi = 0 \rightarrow  \,\, & \frac{\partial^2 \Psi}{\partial q^2} + 12 \pi^4 (\Lambda q - 3)\Psi =0\,. \label{WdWq}
\end{align}
Alternatively, the commutator may be realised by the substitution $q \mapsto \hat{q} =  i  \frac{\partial}{\partial p}$, leading to the Wheeler-DeWitt (WdW) equation in momentum space
\begin{align}
\hat{H}_{(p)} \Psi = 0 \rightarrow  \,\, & (p^2 + 36 \pi^4) \Psi + 12 \pi^4 \Lambda i \frac{\partial \Psi}{\partial p} =0\,, \label{WdWp}
\end{align}
where the only subtlety is that in addition we had to flip the sign $\frac{\partial}{\partial p}\to - \frac{\partial}{\partial p}$ since we are considering this equation at the ``initial'' boundary, as we will discuss presently.

In $4$ dimensions, gravitational path integrals always interpolate between two $3$-dimensional hypersurfaces, on which boundary conditions must be imposed. Moreover, gravitational path integrals satisfy the WdW equation by construction on these two hypersurfaces \cite{Halliwell:1988wc}. It is now crucial to recall the results of \cite{DiTucci:2020weq}: in the case of negative $\Lambda,$ we are interested in calculating the canonical partition function with the $3$-sphere kept fixed at a large radius (in fact, for AdS/CFT this radius is eventually sent to infinity), and with a sum over regular geometries in the interior. Thus at the large radius $q_1$ we would like to impose a Dirichlet condition on the scale factor, and the field space WdW equation \eqref{WdWq} should be satisfied. This explains the inclusion of the York-Gibbons-Hawking boundary term in the action in Eq. \eqref{act}. On the other boundary, however, a regularity condition must be imposed, which corresponds to the imposition of a momentum constraint $p=p_0.$ (Momentum conditions have been repeatedly considered in quantum cosmology, see for instance \cite{Louko:1988bk,Halliwell:1988ik,DiazDorronsoro:2018wro}. The finding of \cite{DiTucci:2020weq} was that one \emph{had} to use a momentum condition in the interior in order to reproduce known results about black hole thermodynamics in asymptotically AdS spacetimes.) Thus on the ``inner'' hypersurface the WdW equation in momentum space \eqref{WdWp} should be satisfied. The specific value that the momentum constraint should take is easy to find: it corresponds to the requirement that the Hamiltonian constraint be satisfied as the geometry caps off at $q=0.$ From Eq. \eqref{Hamiltonian} we see that when $q=0,$ we must demand $p_0^2 + 36 \pi^4 = 0 \rightarrow p_0 = \pm 6\pi^2 i.$ As we will show in section \ref{sec:tensor}, only the choice of sign 
\begin{align}
p_0 = -6\pi^2 i \label{momcond}
\end{align} 
leads to stable perturbations around the most symmetric solution. Note that this implies that on the inner boundary, the wave function satisfies
\begin{align}
\frac{\partial}{\partial p}\Psi  \mid_{inner} = 0\,, \label{noflux}
\end{align}
i.e. we arrive at the interesting notion that there is no momentum flux at the creation of the universe\footnote{This should be contrasted with Vilenkin's tunnelling prescription, according to which the wave function should only contain outgoing modes on the boundaries of superspace \cite{Vilenkin:1986cy}.}.

The considerations above imply that the wave function will factorise
\begin{align}
\Psi(p_0,q_1)=\Psi_{(p)}(p_0)\Psi_{(q)}(q_1),
\end{align}
with $\Psi_{(p)}, \Psi_{(q)}$ satisfying the WdW equation in momentum and position space respectively. Now Eq. \eqref{WdWp} implies that 
\begin{align}
\Psi_{(p)}(p_0)=c_p e^{\frac{3}{\hbar\Lambda}ip_0+\frac{1}{36\pi^4\hbar \Lambda}ip_0^3}\,, \qquad \qquad \Psi_{(p)}(- 6\pi^2 i)=c_p e^{\frac{12\pi^2}{\hbar \Lambda}}\,, 
\end{align}
where we have re-instated $\hbar$ and where $c_p$ is a normalisation constant, which plays no role in what follows and which we will therefore drop. Thus the ``nucleation'' part of the WdW equation is essentially uniquely fixed. 

At the large boundary where $q=q_1,$ the WdW equation \eqref{WdWq} takes the form of an Airy equation. Its general solution is given by
\begin{align}
\Psi_{(q)}(q_1)=c_1 \, Ai\left[ \left(\frac{18\pi^2}{-\hbar\Lambda}\right)^{2/3}\left(1-\frac{\Lambda}{3}q_1\right)\right]+ c_2 \, Bi\left[ \left(\frac{18\pi^2}{-\hbar\Lambda}\right)^{2/3}\left(1-\frac{\Lambda}{3}q_1\right)\right] \label{Adswfgeneralsol}
\end{align}
Here $c_1, c_2$ are two a priori complex integration constants, though once again we do not care about the overall normalisation. Note that for $\Lambda<0,$ the arguments of the Airy functions are real (and positive), and thus the values of the Airy functions are also real. As discussed in \cite{DiTucci:2020weq}, a comparison to the expected CFT result implies that, first of all, the wave function must be real (thus we need $c_1, c_2 \in \mathbb{R}$) and secondly that one expects a volume divergence at large $q_1,$ which will be compensated by counter terms. Given the asymptotic expressions $Ai(x) \sim e^{-\frac{2}{3}x^{3/2}}\,, \quad Bi(x) \sim e^{+\frac{2}{3}x^{3/2}},$ it is clear that the expected result is the $Bi$ function, suggesting that we set $c_1=0.$ We will argue below that this is the correct result, however at this stage it is premature to reach this conclusion, as the $Ai$ function automatically vanishes in the large $q_1$ limit, and thus we cannot exclude its presence in the wave function. This is important in the present context, as we want to look at the analytic properties of the wave function. Moreover, for the Airy functions it is known that small contributions in some parameter ranges can become large (and even dominant) in other parameter ranges, as a reflection of the well known Stokes phenomenon.  

\begin{figure}[h]
	\centering
	\includegraphics[width=0.6\textwidth]{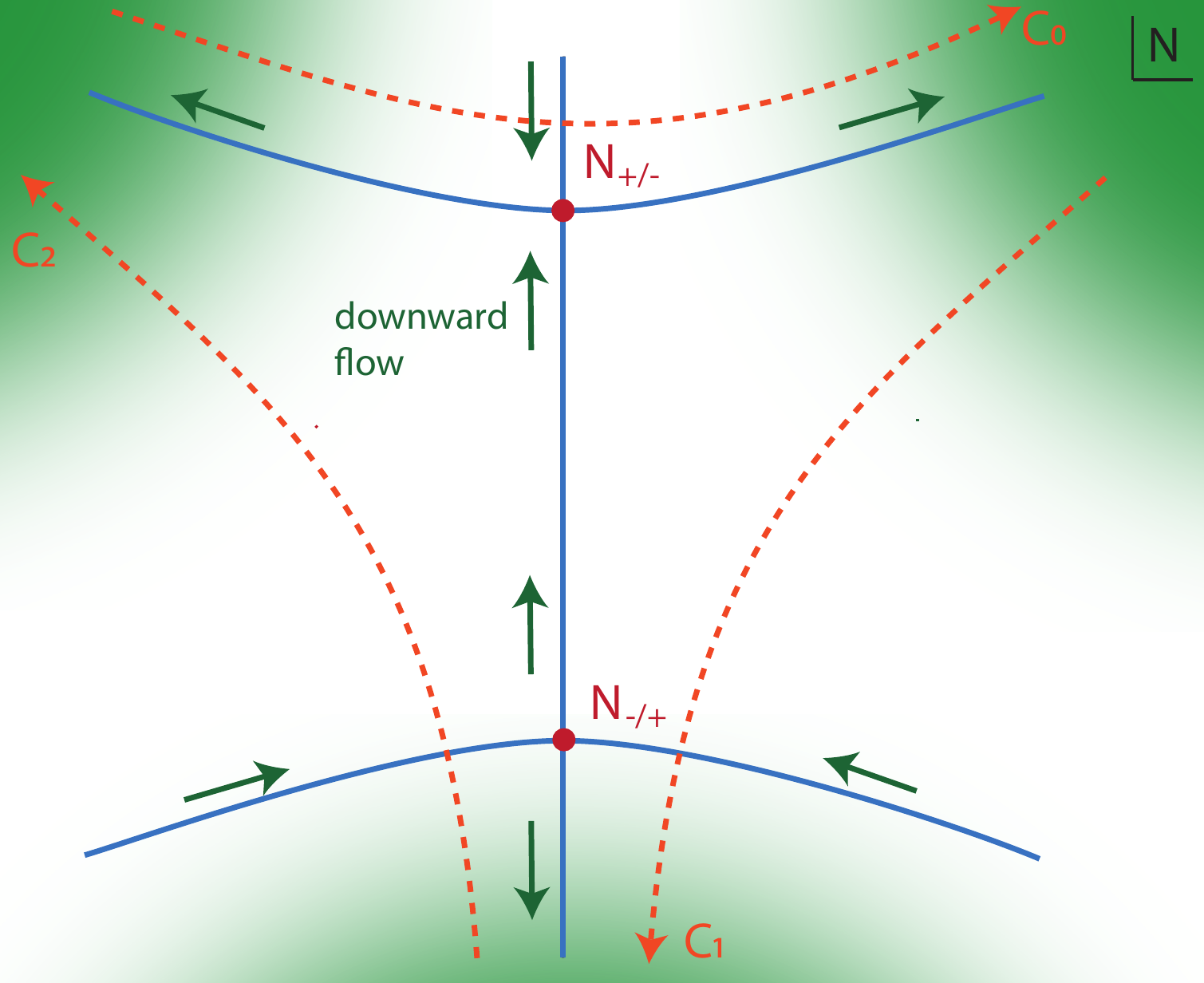}
	\caption{Flow lines for the case of Euclidean saddle points, either for AdS (all $q_1$) or for dS (small universes in the quantum regime, $q_1<\frac{3}{\Lambda}$), in the complexified plane of the lapse $N$. Green regions are regions of asymptotic convergence, at angles $0<\theta<\frac{\pi}{3},$ $\frac{2\pi}{3}<\theta<\pi$ and $\frac{4\pi}{3}<\theta < \frac{5\pi}{3}$. When $\Lambda<0,$ the upper saddle point is $N_+,$ while for $\Lambda>0$ the upper saddle point is $N_-,$ and vice versa for the lower saddle point. The saddle point geometries are shown in Figs. \ref{saddlesneg} and \ref{saddlespos}.}
	\label{PLneg}
\end{figure}

We can make progress by reviewing the path integral representation of the wave function. This calculation was performed in detail in \cite{DiTucci:2019bui,DiTucci:2020weq} (building on \cite{Feldbrugge:2017kzv}) and thus we will not repeat all the steps. In brief: since the Lagrangian \eqref{Lag} is quadratic in the scale factor squared $q,$ we may perform the path integral over $q$ by shifting variables to $q(t)=\bar{q}(t)+Q(t),$ where $\bar{q}$ is a solution of the equation of motion $\ddot{q}=\frac{2}{3}\Lambda N^2$ satisfying the boundary conditions, explicitly
\begin{align}
\bar{q}(t)=\frac{\Lambda}{3}N^2 (t^2 -1) - \frac{p_0}{3\pi^2} N(t -1) + q_1=\frac{\Lambda}{3}N^2 (t^2 -1) +2 N i (t -1) + q_1\,.
\end{align}
The shifted variable $Q(t)$ is an arbitrary perturbation satisfying $\dot{Q}(t=0)=0=Q(t=1).$ Here we have defined the coordinate $t$ such that the inner boundary is at $t=0,$ while the outer boundary is at $t=1.$ (Note that the total time/distance between $t=0$ and $t=1$ will be determined by the value of the lapse $N$.) By construction, the path integral over $Q$ now reduces to a standard Gaussian integral with mixed Neumann-Dirichlet conditions, which just contributes a numerical factor to the wave function (see the appendix of \cite{DiTucci:2020weq}). We are then left with an ordinary integral over the lapse function,
\begin{align}
\Psi(p_0,q_1) & = \int dN e^{i(S_0)/\hbar}\,, \label{lapsepartition}\\
\frac{1}{2\pi^2} S_0(N) &= \frac{\Lambda^2}{9}N^3 - \frac{p_0 \Lambda}{6\pi^2}N^2+\left(-q_1\Lambda + 3 + \frac{p_0^2}{12\pi^4} \right)N+\frac{p_0}{2\pi^2}q_1 \\ 
&= \frac{\Lambda^2}{9}N^3+i\Lambda N^2 - \Lambda q_1 N - 3 q_1 i \,, \label{sphereaction}
\end{align}
which, as one may readily verify, satisfies the WdW equation\footnote{E.g. to check Eq. \eqref{WdWp}, one may use the relation $p_0^2+36\pi^4-12\pi^4 S_{0,p_0}=6\pi^2 S_{0,N},$ and then use the fact that the integration contours over $N$ contain no end points.}. The saddle points of this integral are located at
\begin{align} 
    N_\pm & = \frac{3}{\Lambda}\left(-i \mp \sqrt{\frac{\Lambda}{3}q_1-1} \right)\,, \label{S3saddles} \\
    S_0(N_\pm)     &= \frac{12\pi^2}{\Lambda} \left[-i \pm \left(\frac{\Lambda}{3}q_1-1 \right)^{3/2}\right]\,, \label{saddleactionN}
\end{align}
where we have also included a calculation of the saddle point action. 

\begin{figure}[h]
	\centering
	\includegraphics[width=0.4\textwidth]{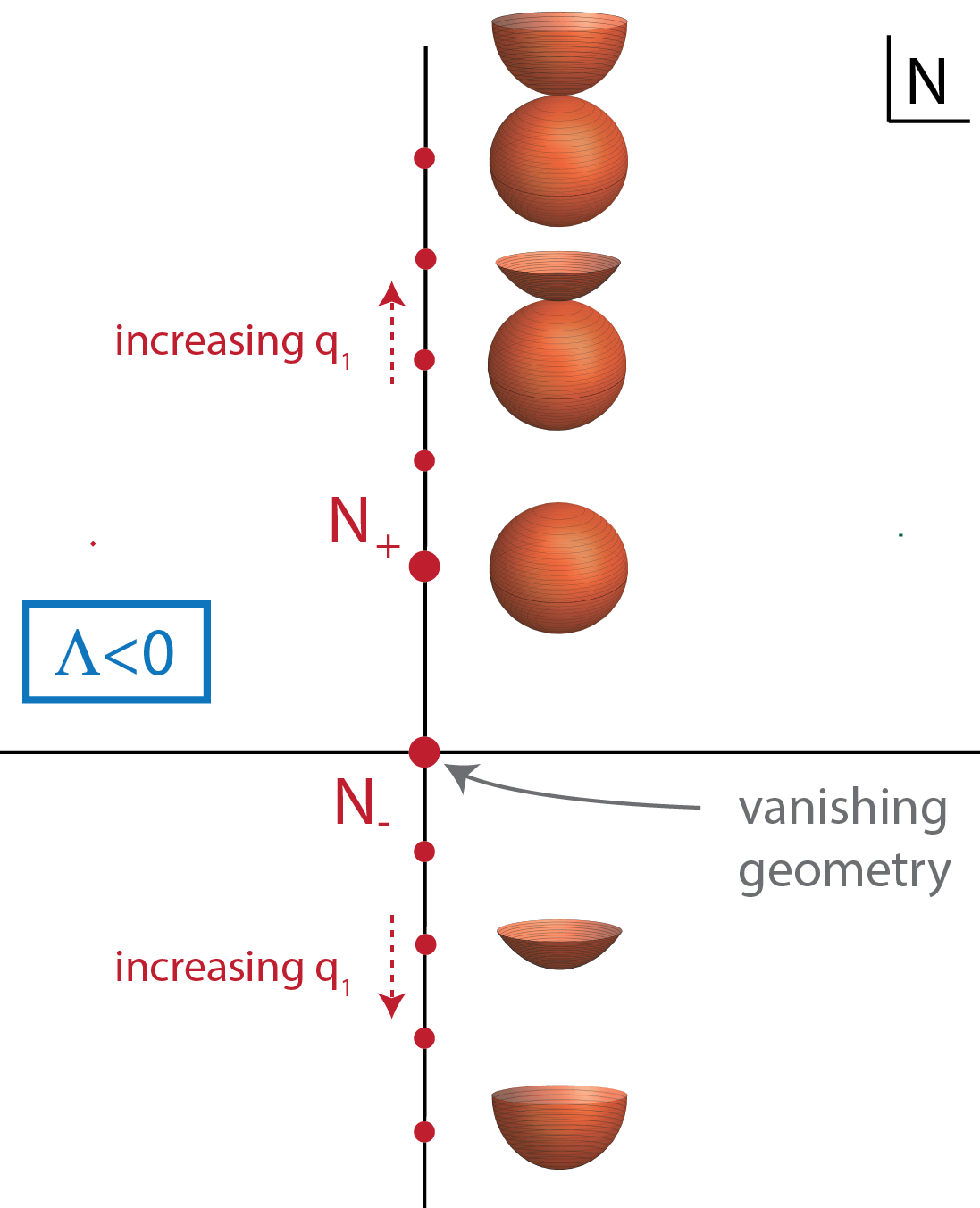}
	\caption{Geometry of Euclidean saddle points, for AdS (all $q_1$). When the outer boundary has zero size, $q_1=0,$ the saddle points are closest together, and in that case $N_-=0$ is associated with a vanishing geometry. As the outer boundary is increased, the saddle points move apart, but remain on the Euclidean axis.}
	\label{saddlesneg}
\end{figure}

The steepest descent contours emanating from the saddle points, as well as the asymptotic regions of convergence, are shown in Fig. \ref{PLneg} for the case where $\Lambda <0$. There are three asymptotic regions of convergence, indicated by the green shaded areas. Thus a priori there are three possible convergent contours of integration, denoted in the figure by ${\mathcal C}_{0,1,2}.$ A comparison to the asymptotic behaviour of the solutions indicates that the integral along ${\mathcal C}_0$ yields the Ai function, while an integration along ${\mathcal C}_2 - {\mathcal C}_1$ yields the Bi function (with an overall factor of $-i$). Thus the Ai part in \eqref{Adswfgeneralsol} stems from the saddle point $N_+,$ while the Bi part comes from $N_-.$ The corresponding geometries are shown in Fig. \ref{saddlesneg}. At the saddle points, the scale factor is given by
\begin{align} \label{saddlemetric}
    \bar{q}(t)\mid_{N_\pm} = \left( q_1-\frac{6}{\Lambda} \pm \frac{6}{\Lambda}i\sqrt{\frac{\Lambda}{3}q_1-1} \right) t^2 + \frac{6}{\Lambda}\left(1 \mp i \sqrt{\frac{\Lambda}{3}q_1 -1}\right) t\,.
\end{align}
When $\Lambda<0,$ the scale factor is purely real. In that case the saddle points are imaginary, resulting in purely Euclidean geometries. Note that even though we imposed a momentum condition at $t=0,$ the scale factor always vanishes at $t=0.$ Thus the saddle point geometries are caping off smoothly, as intended. For $N_-,$ the scale factor only vanishes at $t=0$ and then grows monotonically to the final value $\bar{q}(1)=q_1.$ For $N_+,$ there is a coordinate value $0<t_*<1$ at which the scale factor vanishes a second time. As we will show in section \ref{sec:tensor}, perturbations blow up near this second zero. This indicates that the saddle point $N_+$ is in fact singular, and must be excluded from the sum over histories. This then confirms that the correct result is obtained by summing over the combination of contours ${\mathcal C}_2 - {\mathcal C}_1,$ yielding a partition function that is purely proportional to the Bi function.

\begin{figure}[h]
	\centering
	\includegraphics[width=0.6\textwidth]{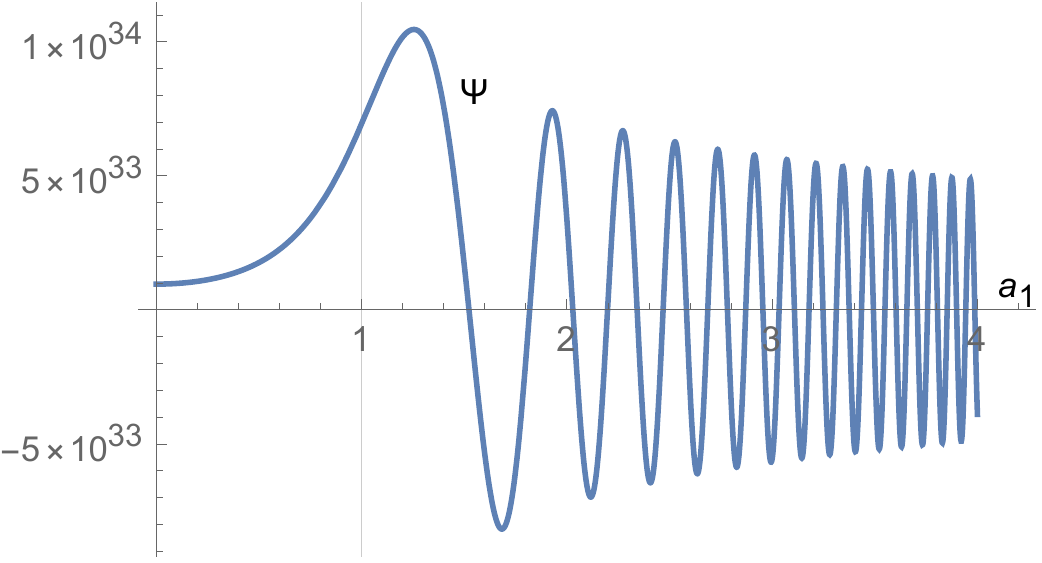}
	\caption{For a universe with spatial 3-spheres, the no-boundary wave function is given by an Airy Ai function. Here the wave function $\Psi$ is plotted as a function of the final scale factor $a_1=\sqrt{q_1}.$ The cosmological constant is set to $\Lambda=1,$ implying that the Stokes phenomenon occurs at $a_1=1.$ For $a_1<1,$ the wave function is approximately exponential (and non-zero at the origin), while for $a_1>1$ it is oscillatory.}
	\label{figAiry}
\end{figure}

After this brief recap, we are in a position to analytically continue the wave function to positive values of the cosmological constant. For this purpose, the following connection formula is most useful,
\begin{align}
& Bi\left[ \left(\frac{18\pi^2}{-\hbar\Lambda}\right)^{2/3}\left(1-\frac{\Lambda}{3}q_1\right)\right] \nonumber \\ = & e^{\frac{\pi}{6}i}Ai\left[ e^{\frac{2\pi}{3}i}\left(\frac{18\pi^2}{-\hbar\Lambda}\right)^{2/3}\left(1-\frac{\Lambda}{3}q_1\right)\right] + e^{-\frac{\pi}{6}i}Ai\left[ e^{-\frac{2\pi}{3}i}\left(\frac{18\pi^2}{-\hbar\Lambda}\right)^{2/3}\left(1-\frac{\Lambda}{3}q_1\right)\right] \\ = & \sqrt{3} \, Ai\left[ \left(\frac{18\pi^2}{\hbar\Lambda}\right)^{2/3}\left(1-\frac{\Lambda}{3}q_1\right)\right]\,.
\end{align}
Thus, for $\Lambda>0,$ the wave function is once again real, but now it is better thought of as given by the Ai function, rather than the Bi function. Its full expression, up to normalisation, is given by
\begin{align}
\Psi(p_0,q_1) = e^{\frac{3}{\hbar\Lambda}ip_0+\frac{1}{36\pi^4\hbar \Lambda}ip_0^3}\, Ai\left[ \left(\frac{18\pi^2}{\hbar\Lambda}\right)^{2/3}\left(1-\frac{\Lambda}{3}q_1\right)\right]\,.
\end{align}
It is worth emphasising that the wave function is thus uniquely defined. A graphical representation is provided in Fig. \ref{figAiry}, where one may immediately see its main characteristics: it rises exponentially from $q_1=0$ (though note that $\Psi(q_1=0)\neq 0$) up to about $q_1=\frac{3}{\Lambda},$ and for larger values of $q_1$ it oscillates with increasing frequency and slightly diminishing amplitude.

\begin{figure}[h]
	\centering
	\includegraphics[width=0.7\textwidth]{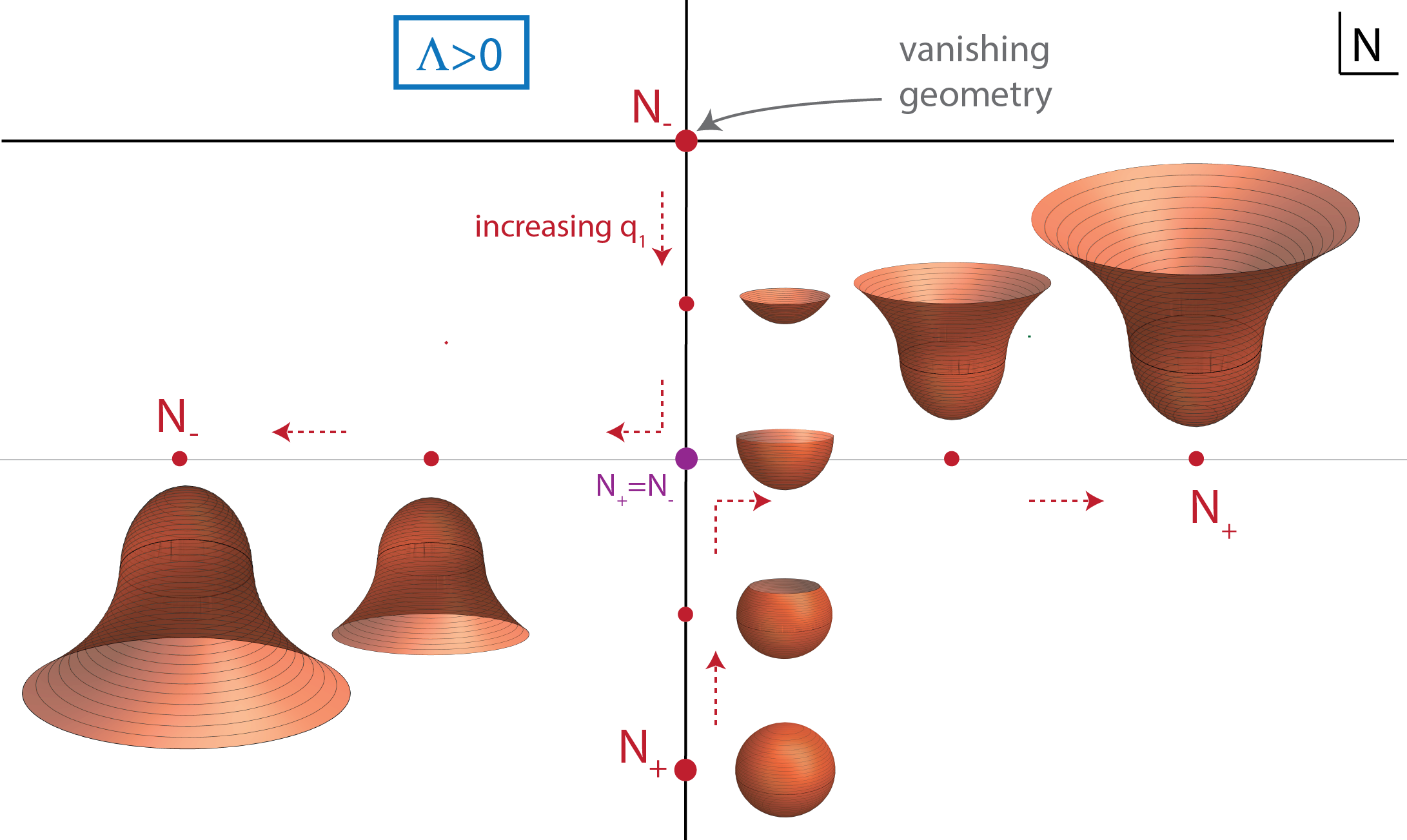}
	\caption{Geometry of saddle points, for dS. For $q_1 \leq \frac{3}{\Lambda}$ the saddle point geometries are Euclidean and correspond to portions of the 4-sphere, while for $q_1 > \frac{3}{\Lambda}$ they are complex, with increasingly Lorentzian (dS) asymptotics.}
	\label{saddlespos}
\end{figure}

It is again illustrative to see what the implications of the exact solution are for the path integral representation. When the final radius of the universe is small, $q_1 < \frac{3}{\Lambda},$ the saddle points are at imaginary values of the lapse, and the geometries are Euclidean. This is very much like the situation for $\Lambda <0,$ cf. Fig. \ref{PLneg}. However, compared to the AdS case, the contour of integration now must be different, so as to yield the Ai function. More precisely, the contour of integration for the lapse integral must be the contour denoted ${\mathcal C}_0$ in that figure. Note that this is implied by the analytic continuation. This means that, in contrast to the AdS case, it is now the upper saddle point that is relevant to the path integral, and not the lower one. The geometry at the saddle points is shown graphically in Fig. \ref{saddlespos}. At small $q_1,$ the geometries are Euclidean and consist of portions of the 4-sphere, with $N_-$ containing less than an hemisphere, and $N_+$ containing more than one hemisphere. At the nucleation of the universe, i.e. for $q_1=0,$ the relevant saddle point $N_-$ corresponds to a vanishing geometry, while the irrelevant saddle at $N_+$ corresponds to a full sphere.

\begin{figure}[h]
	\centering
	\includegraphics[width=0.6\textwidth]{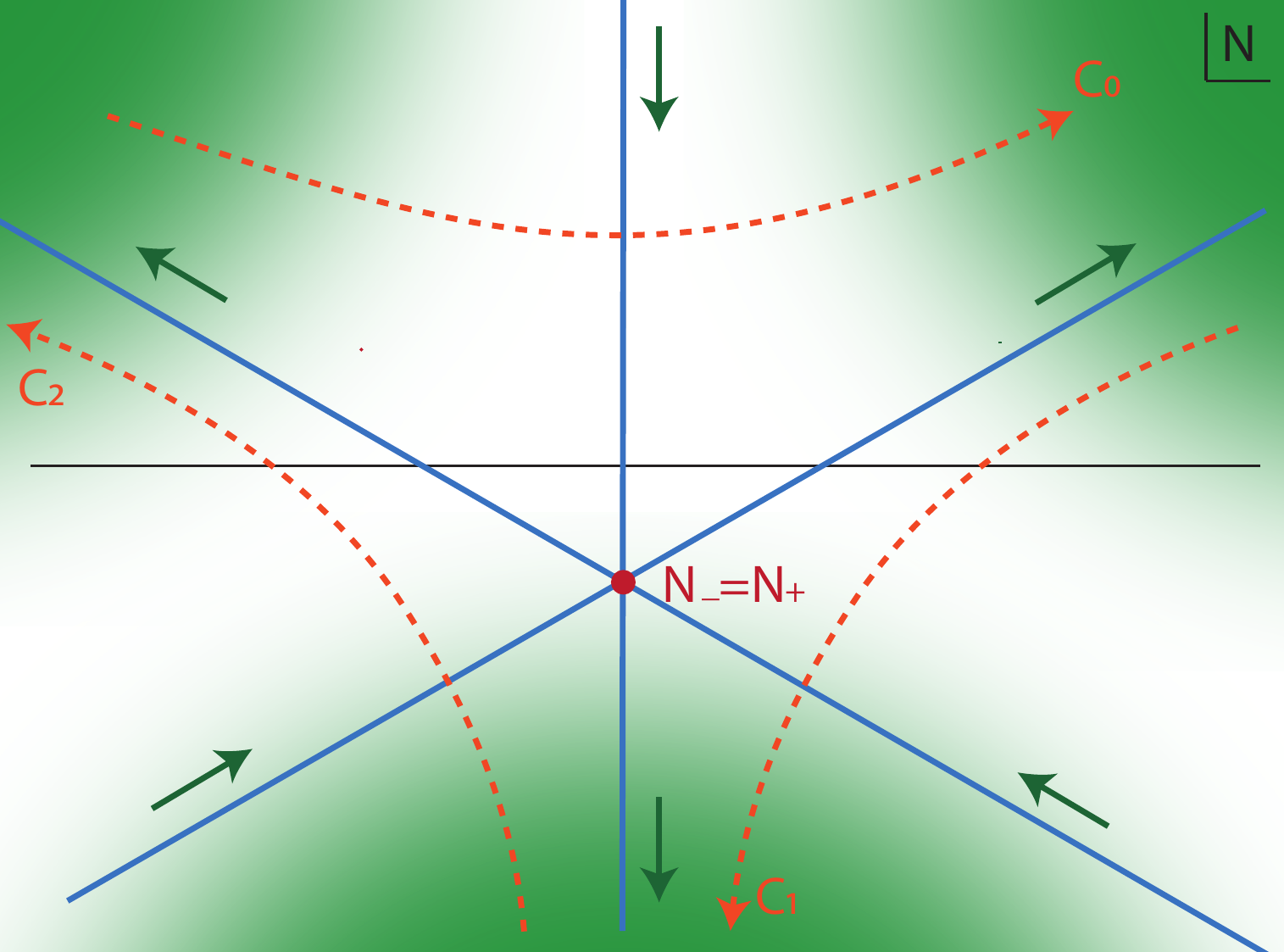}
	\caption{Flow lines for the case where the saddle points degenerate, i.e. $\Lambda>0$ with $q_1=\frac{3}{\Lambda}$. This graph represents the cross-over between Figs. \ref{PLneg} and \ref{PLpos}.}
	\label{PLdegenerate}
\end{figure}

As the universe grows, in contrast to the AdS case, the saddle points now approach each other and coalesce when $q_1=\frac{3}{\Lambda}.$ The steepest descent lines for the resulting degenerate saddle (sometimes called a monkey saddle) are shown in Fig. \ref{PLdegenerate}. The geometry at the saddle point is that of exactly one hemisphere. Up to this radius, the wave function behaves approximately exponentially,
\begin{align}
\Psi(p_0,q_1) \approx e^{\frac{3}{\hbar\Lambda}ip_0+\frac{1}{36\pi^4\hbar \Lambda}ip_0^3}\, e^{-\frac{12\pi^2}{\hbar \Lambda}(1-\frac{\Lambda}{3}q_1)^{3/2}}=e^{\frac{12\pi^2}{\hbar \Lambda}\left[1-(1-\frac{\Lambda}{3}q_1)^{3/2}\right]}\,, \quad\qquad q_1 \leq \frac{3}{\Lambda}\,.
\end{align}
Note that the weighting increases until $q_1=\frac{3}{\Lambda}$ is reached.

\begin{figure}[h]
	\centering
	\includegraphics[width=0.6\textwidth]{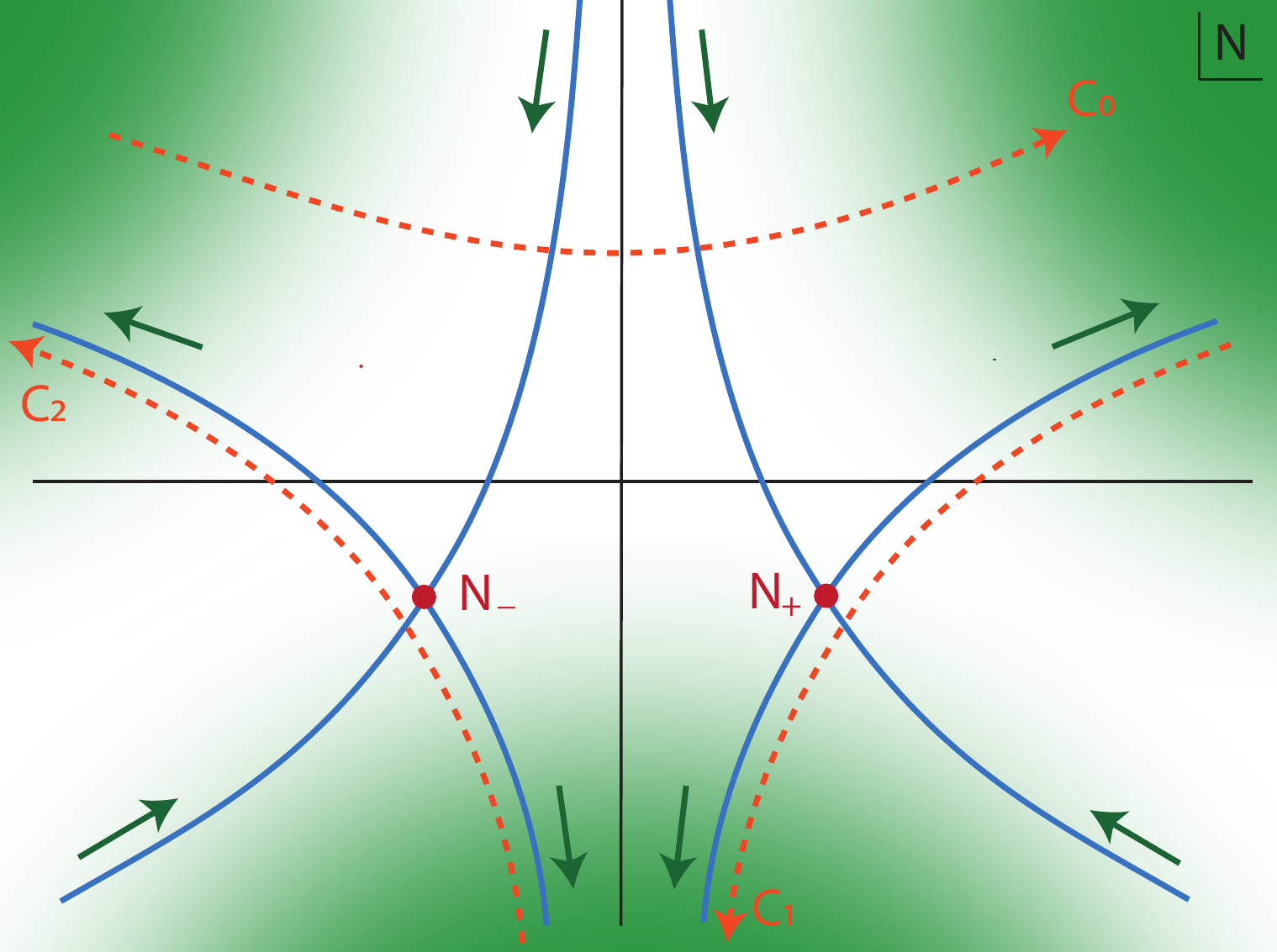}
	\caption{Flow lines for the case of complex saddle points with Lorentzian asymptotics, i.e. for large universes with $q_1>\frac{3}{\Lambda},$ with $\Lambda>0$.}
	\label{PLpos}
\end{figure}

As the universe grows further, $q_1> \frac{3}{\Lambda},$ the degenerate saddle separates into two saddles which are both relevant to the path integral, see Fig. \ref{PLpos}. The contour of integration for the lapse function remains ${\mathcal C}_0 = - {\mathcal C}_2 - {\mathcal C}_1.$  The saddle points are now located at complex values of the lapse, with a real part that increases in magnitude as the universe grows. The two saddle point geometries are of no-boundary type, meaning that they start at $q(0)=0$ and have an asymptotic region near $t=1$ that is ever closer to Lorentzian dS spacetime as $q_1$ grows. The geometries are complex conjugates of each other. The transition from one relevant saddle to two relevant saddles is known as a Stokes phenomenon. It has important implications for the physical meaning of the wave function. In particular, the wave function changes from exponential to oscillatory in this regime, with its approximate expression given by 
\begin{align}
\Psi(p_0,q_1) &\approx \frac{1}{2i}e^{\frac{12\pi^2}{\hbar \Lambda}}\, \left[e^{i\frac{12\pi^2}{\hbar \Lambda}(\frac{\Lambda}{3}q_1-1)^{3/2}+i\frac{\pi}{4}}-e^{-i\frac{12\pi^2}{\hbar \Lambda}(\frac{\Lambda}{3}q_1-1)^{3/2}-i\frac{\pi}{4}}\right] \,, \quad\qquad q_1 > \frac{3}{\Lambda} \\ &=e^{\frac{12\pi^2}{\hbar \Lambda}}\,\sin\left[\frac{12\pi^2}{\hbar \Lambda}(\frac{\Lambda}{3}q_1-1)^{3/2} +\frac{\pi}{4}\right] \,, \quad\qquad q_1 > \frac{3}{\Lambda} \,.
\end{align}
The fact that the wave function oscillates ever faster with increasing $q_1$ while its weighting remains constant implies that the wave function becomes classical in the WKB sense. Therefore, successive path integrals with increasing real boundary values $q_1$ describe real Lorentzian dS universes (even though the saddle points of each individual path integral have a complex geometry), as long as $q_1 > \frac{3}{\Lambda}$. 

We should highlight that time only appears after a Stokes phenomenon has occurred. This is a direct consequence of the definition of the path integral: it is defined with a Euclidean momentum condition at the nucleation of the universe. Thus the universe necessarily starts off Euclidean, and there is no time yet. In order for time to start existing, the wave function must become oscillatory. Thus the relevant saddle point(s) has/have to become complex. But since the wave function is by definition real (as implied by the analytic continuation from AdS), the wave function can only be oscillatory if it is composed of two complex conjugate saddles. Thus we may identify a link between the emergence of time from space, and a Stokes phenomenon in the path integral defining the wave function of the universe. From this point of view it is then inevitable that for any quasi-classical geometry contributing  to the wave function, its complex conjugate (and thus time reversed) geometry will also contribute. 

A further remark is in order:  the irrelevant saddle point $N_+$, for $q_1=0,$ has scale factor given by $\bar{q}=\frac{12}{\Lambda}(t-t^2)$ and corresponds to a full 4-sphere. If we had added a scalar field, then the natural boundary condition would have been a Dirichlet boundary condition for the scalar. At a maximum of the scalar potential one would then have obtained the Hawking-Moss instanton \cite{Hawking:1981fz}, with constant scalar field. At other (sufficiently flat) locations on the potential the geometry would have been a slightly deformed 4-sphere with almost constant scalar and, for potentials with an appropriate barrier, the instantons would have included those of Coleman-De Luccia type \cite{Coleman:1980aw}, with the scalar interpolating across the barrier\footnote{Our momentum condition \eqref{momcond} implies that these instantons would in fact be the complex conjugates of the usual HM and CdL instantons considered in tunnelling phenomena. Their geometries would be identical.}. The point is now that all these instantons do not contribute to the nucleation of the universe, according to the no-boundary wave function (as defined via analytic continuation from AdS path integrals). This is perhaps surprising: as we saw above, the no-boundary wave function does not vanish when the size of the universe is set to zero, i.e. $\Psi(q_1=0)\neq 0.$ This property was already noticed by Hartle and Hawking, who attributed this fact to contributions from non-trivial topologies \cite{Hartle:1983ai}. The reason for this interpretation is that topology change occurs precisely at zero scale factor (in an appropriate slicing). Hence a diagnostic for the contribution of non-trivial topologies would be a non-vanishing wave function at zero scale factor. The simplest non-trivial topologies are of 4-sphere form (and at larger scale factors, one can think of a 4-sphere being glued onto the bottom of the geometry with trivial topology), i.e. of HM or CdL type. Here we see that such geometries however do not contribute, casting doubt on the original interpretation of Hartle and Hawking. To reinforce this point, note that the action of the 4-sphere would give a contribution proportional to $e^{24\pi^2/(\hbar \Lambda)},$ cf. Eq. \eqref{saddleactionN}. This would vastly dominate over the contribution from the vanishing geometry, and similar considerations also apply when $0 < q_1 < \frac{3}{\Lambda}.$ Thus, if they were not eliminated outright, the simplest non-trivial topologies would dominate the wave function. Maybe a better interpretation stems from considering the definition of the path integral as a sum over all regular geometries: the uncertainty principle implies that since we have fixed the initial momentum, the sum over histories must include a sum over universes with all possible initial sizes. Thus, even when the universe is tiny, it may still receive some small contributions from geometries of vasty different sizes. This point seems worthy of further exploration in the future.

\section{Adding Tensor Perturbations} \label{sec:tensor}

In order to assess the cosmological implications of the wave function of the universe, one must add fluctuations. This is doubly important, as fluctuations provide a link to the early universe via the CMB, but on a purely theoretical level also, it has been realised over the last few years that a quantisation of background plus fluctuations may yield non-trivial results: in particular, certain backgrounds may lead to unsuppressed fluctuations \cite{DiTucci:2019xcr} -- this is how it was discovered that a definition of a no-boundary path integral with purely Dirichlet boundary conditions is not tenable \cite{Feldbrugge:2017fcc}. The first calculations of no-boundary fluctuations already date back to the 1980s \cite{Halliwell:1984eu}, hence we do not need to repeat all results. Rather, we will illustrate no-boundary fluctuations from the perspective of the minisuperspace path integral discussed in the previous section, which will reveal a few hitherto little discussed aspects of these fluctuations, and relate them to fluctuations in AdS spaces. 

Our analysis will proceed at the level of the saddle point approximation to the path integral, neglecting the backreaction of the perturbations on the background. At the saddle points, the geometries are given by
\begin{align}
\bar{q}_\pm(t) = \frac{\Lambda}{3}N_\pm^2t^2 + 2 N_\pm i t\,.
\end{align}
Tensor perturbations then obey the equation of motion \cite{Feldbrugge:2017fcc}
\begin{align}
\ddot{h} + 2 \frac{\dot{q}}{q}\dot{h} + \frac{N_\pm^2}{q^2}k(k+2)h=0\,, 
\end{align}
where $h(t)$ denotes a Fourier component of a fluctuation mode expanded into spherical harmonics with wave number $k.$ Solutions with $h(t=1)=h_1$ are given by $h(t)=h_1 \frac{F(t)}{F(1)}$ with the time/space dependence encoded in \cite{Feldbrugge:2017fcc}
\begin{align}
F(t) & = \left(1-\frac{i}{\frac{\Lambda}{3}N_\pm t+i} \right)^{\frac{k}{2}}\left(1+\frac{i}{\frac{\Lambda}{3}N_\pm t+i} \right)^{-\frac{(k+2)}{2}}\left(1+\frac{i(k+1)}{\frac{\Lambda}{3}N_\pm t+i} \right) \label{pertst}\\
& = \frac{\left( \frac{\Lambda}{3}N_\pm t \right)^{\frac{k}{2}}\left[\frac{\Lambda}{3}N_\pm t+(k+2)i \right]}{\left( \frac{\Lambda}{3}N_\pm t+2i\right)^{\frac{k+2}{2}}}\,. \label{pertscompact}
\end{align} 
Here we have picked the solution for which $F(0)=0,$ ensuring regularity at $t=0.$ A second solution exists, but this solution diverges at $t=0,$ and is thus unphysical.

\begin{figure}[h]
	\centering
	\includegraphics[width=0.45\textwidth]{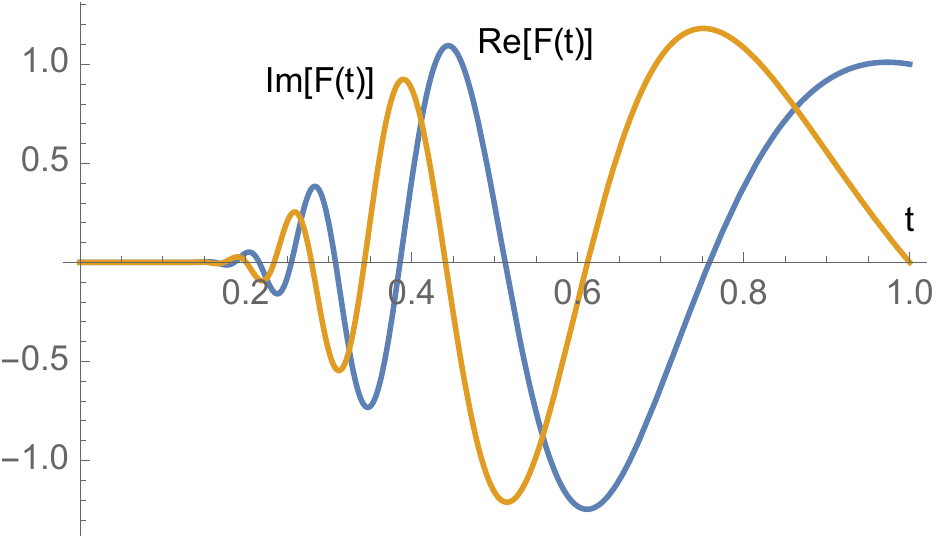}
	\caption{Normalised tensor perturbations at the relevant no-boundary saddle point, with $\Lambda=1,$ $q_1=1201$ and for wave number $k=100.$}
	\label{PertsdS}
\end{figure}

At the saddle point $N_-,$ the tensor perturbations are well behaved. They give the action \cite{Feldbrugge:2017fcc}
\begin{align}
S_h^{(k)} &=\frac{1}{2}\int dt\, N \left[ q^2\left( \frac{\dot{h}}{N}\right)^2 - k(k+2)h^2\right] \\ &= \frac{h_1^2}{2}\left[ -k(k+2)\sqrt{\frac{3q_1}{\Lambda}}+ i \frac{3k(k+1)(k+2)}{2\Lambda} + {\mathcal O}(\frac{1}{\sqrt{q_1}})\right]
\end{align}
For $\Lambda>0,$ they yield a suppressed Gaussian distribution $\Psi^{(k)} \propto e^{-\frac{3k(k+1)(k+2)}{4\hbar\Lambda}h_1^2}$ and correspond to the Bunch-Davies vacuum in the closed slicing of de Sitter space, see Fig.~\ref{PertsdS} for a graphical representation. Had we chosen the opposite momentum condition in Eq. \eqref{momcond}, we would have ended up with an inverse Gaussian distribution, and consequently an ill-defined model \cite{Feldbrugge:2017fcc}. For $\Lambda<0,$ the wave function is pure imaginary, and the divergence at large volume would have to be cancelled by a counter term, as usual \cite{Maldacena:2002vr}. 

\begin{figure}[h]
	\centering
	\includegraphics[width=0.45\textwidth]{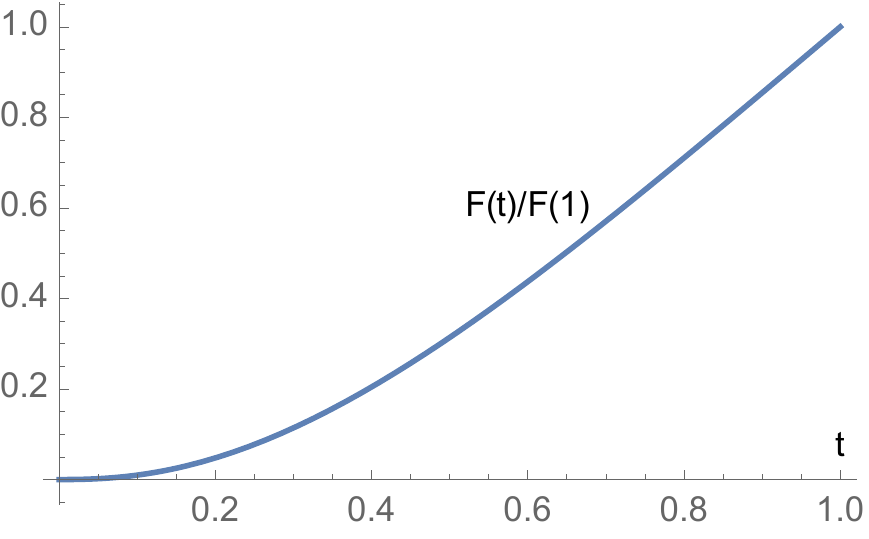}
	\includegraphics[width=0.45\textwidth]{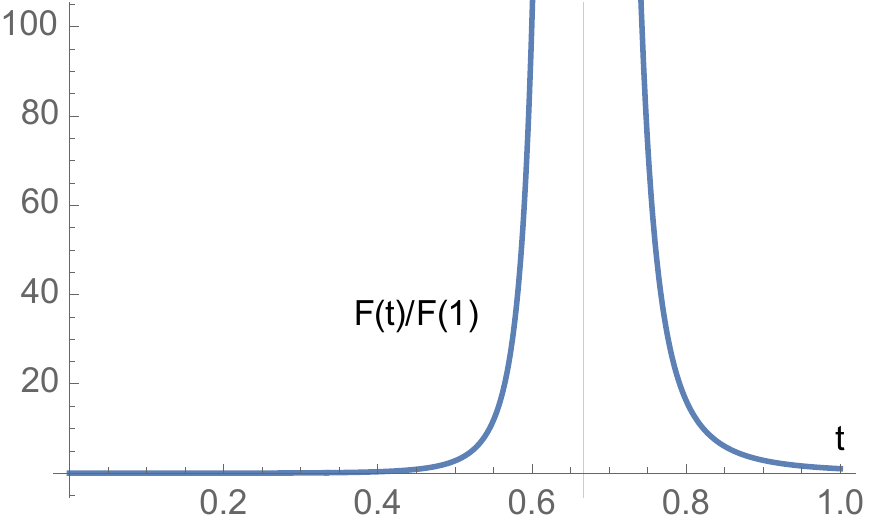}
	\caption{Tensor perturbations $h(t)=h_1 \frac{F(t)}{F(1)}$ for AdS saddle point geometries, for $k=5$. The left panel shows the perturbation at the regular saddle point $N_-.$ Note that in the case of a negative $\Lambda<0,$ the perturbations grow from zero to their final size without any oscillations. The right panel shows the perturbation at the saddle point $N_+,$ where the scale factor passes through zero (vertical thin grey line), and where consequently the perturbation blows up.}
	\label{PertsAdS}
\end{figure}

At the saddle point $N_+,$ there is always a value $0 < t_* \leq 1$ at which $q_+(t_*)=0.$ At that location, the tensor perturbations blow up, i.e. $F(t) \to \infty$ as $t\to t_*.$ A numerical example is shown in Fig.~\ref{PertsAdS} for $\Lambda < 0.$ This confirms that the saddle points $N_+$ are not physical, and should not be included in path integral. This restricts the possible contours of integration for the lapse function, and lead to the selection of  the $Bi$ solution only in \eqref{Adswfgeneralsol}. 

It is useful to also present the expression for the perturbations in physical coordinates, with the metric $ds^2 = - dt_p^2 + a^2 d\Omega_3^2.$ The relation between the $t$ coordinate and $t_p$ is given by
\begin{align}
\sinh \left( \sqrt{\frac{\Lambda}{3}}t_p\right) = \frac{\Lambda}{3}Nt+i\,, \label{complextime}
\end{align}
so that the solution for the perturbations \eqref{pertst} becomes
\begin{align}
F(t_p)  &= \left(1-\frac{i}{\sinh \left( \sqrt{\frac{\Lambda}{3}}t_p\right) } \right)^{\frac{k}{2}}\left(1+\frac{i}{\sinh \left( \sqrt{\frac{\Lambda}{3}}t_p\right) } \right)^{-\frac{(k+2)}{2}}\left(1+\frac{i(k+1)}{\sinh \left( \sqrt{\frac{\Lambda}{3}}t_p\right) } \right) \\ &= \left(1-\frac{1}{\sin \left( \sqrt{\frac{-\Lambda}{3}}t_p\right) } \right)^{\frac{k}{2}}\left(1+\frac{1}{\sin \left( \sqrt{\frac{-\Lambda}{3}}t_p\right) } \right)^{-\frac{(k+2)}{2}}\left(1+\frac{(k+1)}{\sin \left( \sqrt{\frac{-\Lambda}{3}}t_p\right) } \right)\,,
\end{align} 
where the top line is more convenient for dS and the bottom expression more convenient for AdS. These expressions are the closed slicing analogues of the much more familiar expressions for perturbations in the flat slicing\footnote{That is to say that for dS they are the analogue of the standard solution $F(\eta)=e^{ik\eta}(1-ik\eta),$ expressed here in terms of conformal time $\eta=-1/(Ha)=-\sqrt{\frac{3}{\Lambda}}e^{-\sqrt{\frac{\Lambda}{3}}t_p}.$ See \cite{Maldacena:2002vr} for a discussion of the relation between dS and AdS perturbations in the flat slicing.}. The inner boundary ($t=0$) is located at $t_p=i\sqrt{\frac{3}{\Lambda}}\frac{\pi}{2}=\sqrt{\frac{3}{-\Lambda}}\frac{\pi}{2},$ and the perturbations vanish there, as required for regularity. 

\begin{figure}[h]
	\centering	
	\includegraphics[width=0.31\textwidth]{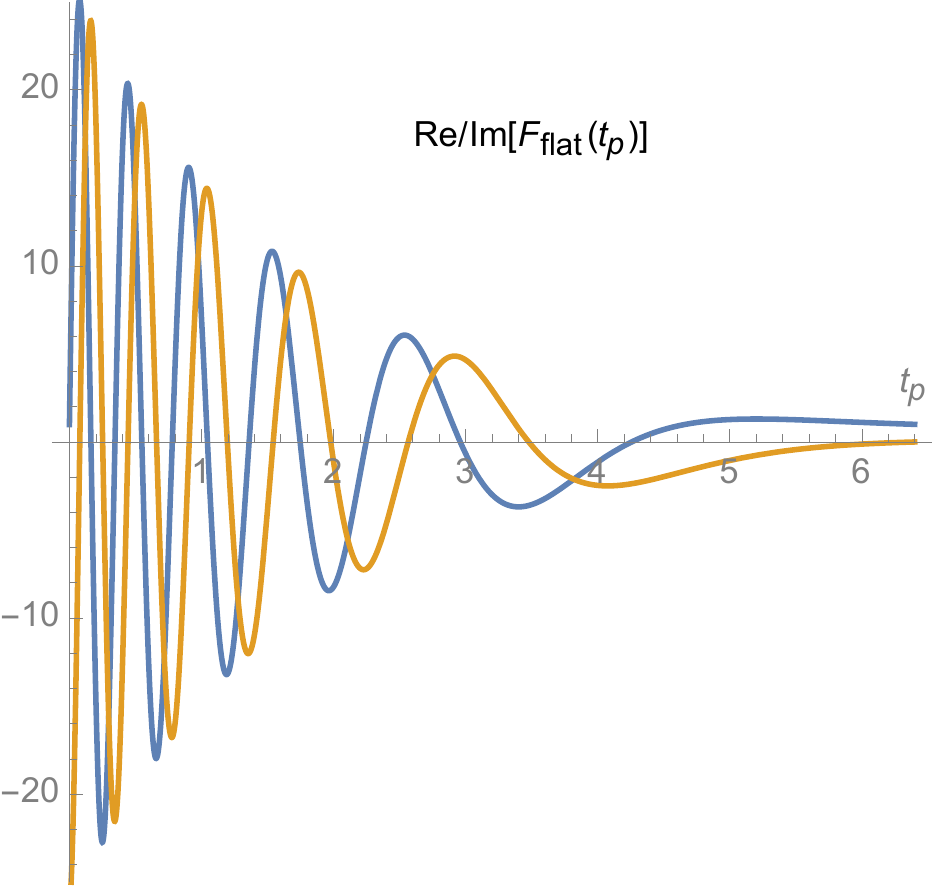}
	\includegraphics[width=0.31\textwidth]{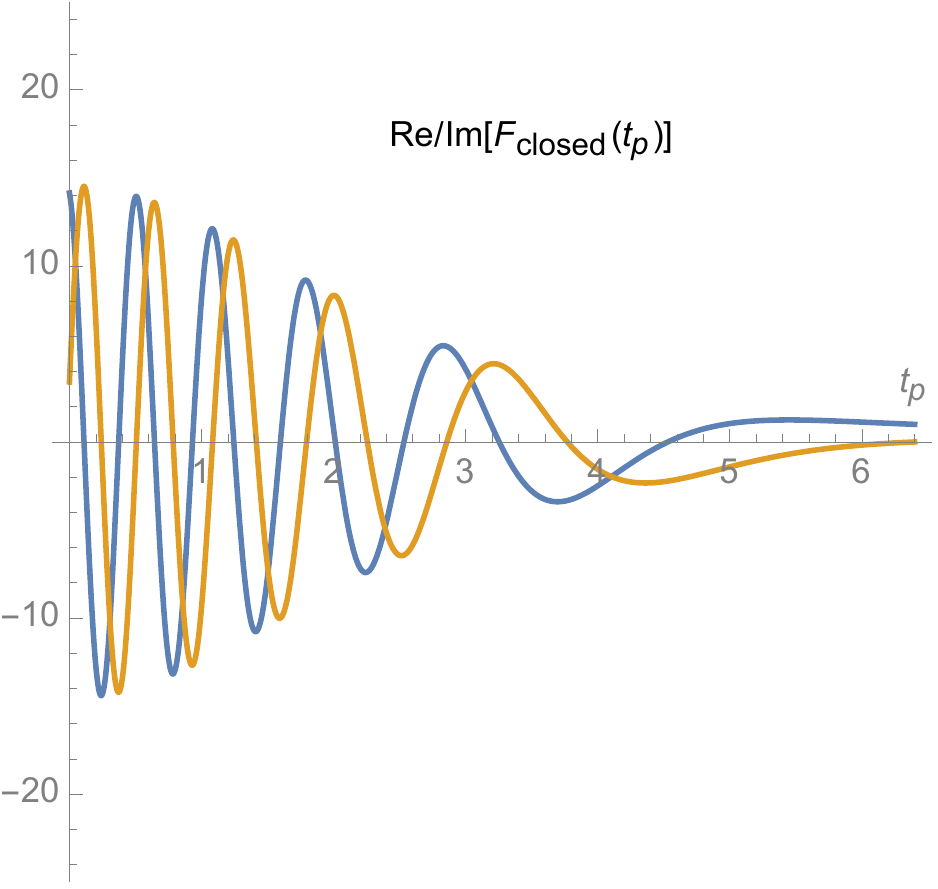}
	\includegraphics[width=0.31\textwidth]{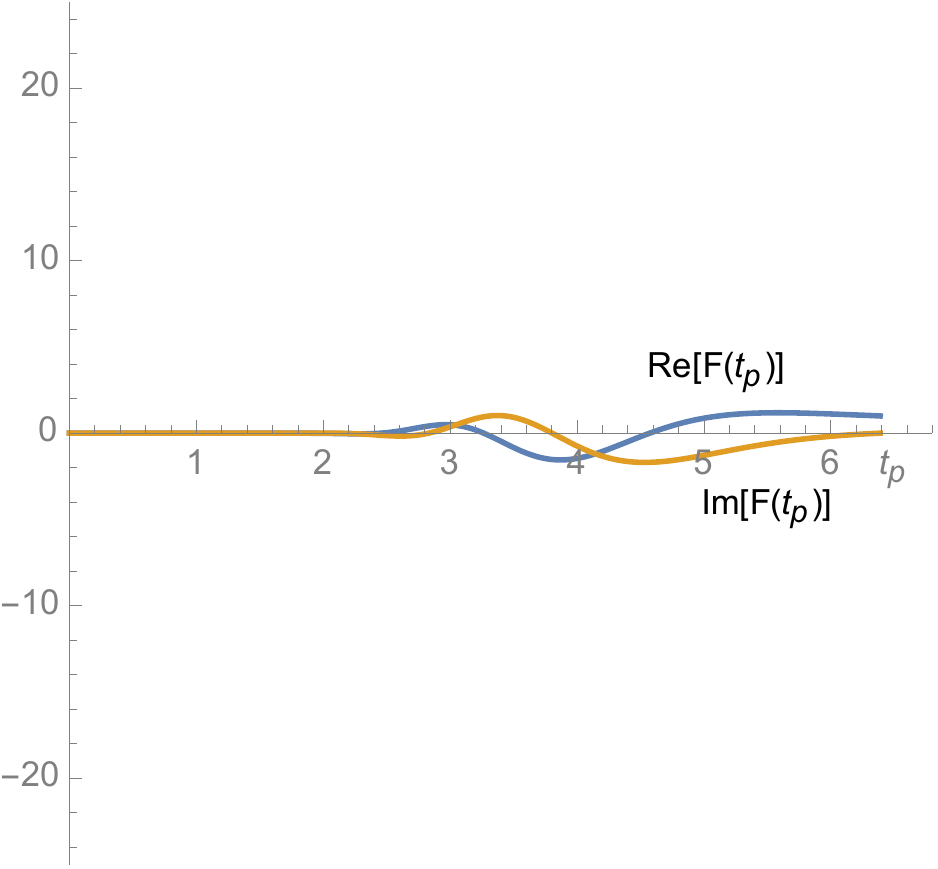}
	\caption{Tensor perturbations $h(t_p)=h_1 F(t_p)$ for dS geometries, for $k=20$. The left panel shows the perturbations for the flat slicing most often employed in inflationary calculations. The centre panel shows the perturbations in real physical time for the closed slicing of dS spacetime, and the right panel at the no-boundary saddle point $N_-,$ but re-scaled to ``physical'' time $t_p.$ The function $F$ is normalised such that $F=1$ at final times. Real parts are shown in blue, and imaginary parts in orange. Here we used $\Lambda=1, q_1=1201.$}
	\label{PertsDiffTimes}
\end{figure}

One can see that for $\Lambda<0,$ the range of $t_p$ will be over real values, while for $\Lambda>0$ we must allow $t_p$ to take on complex values. The range $0 \leq t \leq 1$ is then translated into a path in the complex $t_p$ plane. This has significant implications for the shape of the perturbations. Fig.~\ref{PertsDiffTimes} shows the same perturbation mode, namely a mode with wave number $k=20,$ on different representations/slicings of (complexified) de Sitter spacetime, each time as a function of the coordinate $t_p$. The time evolution is shown up to the point of horizon crossing, where the mode freezes. The left panel shows the evolution of a tensor perturbation, normalised such that at the end it equals unity, in the flat slicing. This is familiar from standard inflationary calculations. In the past the perturbation oscillates rapidly. The central panel shows the same perturbation, but now on a closed slicing of dS, where the time evolution starts at the waist (minimum radius) of the dS hyperboloid. The perturbation oscillates equally rapidly as in the flat slicing, but the amplitude is somewhat reduced near the minimum size of dS. The right panel shows the same perturbation mode, but now along the complex path in the $t_p$ plane implied by Eq. \eqref{complextime}, for the relevant saddle point $N_-$. The difference is striking: the oscillations are far fewer in number, the overall amplitude is significantly reduced, and at early times the amplitude is very significantly damped. We can understand this from an inspection of Eq. \eqref{pertscompact}, which shows that for small $t,$ the perturbations are suppressed as $t^{k/2}.$ No-boundary geometries are often represented by a Euclidean half-sphere glued onto half of a Lorentzian dS hyperboloid\footnote{That this representation cannot be the appropriate one is already indicated by studies of no-boundary saddle points with large anisotropies \cite{Bramberger:2017rbv}.}. Had we used such a representation, we would not have seen the smoothing effect on the perturbations: going back in time, we would have seen the evolution such as that in the centre panel of Fig. \ref{PertsDiffTimes} first, and then in the Euclidean part the perturbations would have shrunk to zero. Here we have used the complexified dS representation given to us by the minisuperspace path integral, and this exhibits the full smoothing effect of the no-boundary condition. This provides a slightly new take on the much-discussed trans-Planckian issue of perturbations \cite{Bedroya:2019snp,Dvali:2020cgt}: we see that high wave number modes are very strongly suppressed when the universe is small, as $F \propto t^{k/2}.$ Moreover they do not oscillate at all until the universe has grown to larger than $q_1 \approx \frac{3}{\Lambda}.$ Hence one might question whether there is any operational meaning to testing such perturbations in the near- or trans-Planckian regime. Certainly, the quantum-to-classical transition of both background and perturbations deserves further study in this context.

\section{Scalar Field} \label{sec:scalar}

The previous sections have shown that one may analytically continue the wave function of the universe from negative to positive cosmological constant, and that this correspondence can be extended to gravitational perturbations of the geometries. This raises the question as to how general such a correspondence might be. As a first step in trying to answer that question, we will consider the case of gravity coupled to a scalar field, i.e. we will consider the action
\begin{align}
S = \frac{1}{2}\int d^4 x \sqrt{-g}\left[ R - (\partial\phi)^2 - 2V(\phi)\right]\,.
\end{align}
We will use a potential of the form of a mass term, $V(\phi)=\frac{1}{2}m^2 \phi^2 = - \frac{1}{2}\mu^2 \phi^2,$ where we will allow the potential to be either positive or negative/tachyonic. It is not known how to solve the path integral analytically in such a model, even in minisuperspace, hence we will proceed at the level of the saddle point approximation of the path integral.

We will first analyse the case of a negative potential, i.e. we will assume $\mu$ to be real. Since we can only work at the level of the saddle point approximation, there is no reason to re-scale the time coordinate as in \eqref{metric}, but it will prove useful to use Euclidean time $\tau = i t_{physical}.$ Thus our ansatz for the background evolution is 
\begin{align}
ds^2 = d\tau^2 + a(\tau)^2 d\Omega_3^2\,, \quad \phi = \phi(\tau)\,.
\end{align}
The equations of motion and the constraint are given by 
\begin{align}
\phi_{,\tau\tau} + 3 \frac{a_{,\tau}}{a}\phi_{,\tau}+ \mu^2 \phi &=0\,, \\
3a_{,\tau\tau}+ a \phi_{,\tau}^2-\frac{1}{2}a \mu^2 \phi^2 &=0\,, \\
3a_{,\tau}^2 - \frac{1}{2}a^2 \phi_{,\tau}^2 -3 - \frac{1}{2}a^2 \mu^2\phi^2 & = 0\,.
\end{align}
We would like to find a solution that is regular as the geometry caps off, i.e. that is regular at $a=0$ (where we will choose the origin of the time coordinate such that $a(\tau=0)=0$). The constraint shows that this requires $a_{,\tau}(\tau=0)=\pm 1,$ and the requirement that fluctuations be stable fixes the choice of sign to $a_{,\tau}(\tau=0)=+1.$ A Taylor expansion of the equations of motion then shows that one should also specify $\phi_{,\tau}(\tau=0)=0,$ while the value of $\phi$ itself remains free; we will denote it by $\phi(\tau=0)=\phi_0.$ An approximate solution near the origin is then given by
\begin{align}
\phi \approx \phi_0\,, \qquad a \approx \frac{\sinh (\frac{\mu}{\sqrt{6}} \phi_0 \tau)}{\frac{\mu}{\sqrt{6}} \phi_0}\,. \label{approxsmall}
\end{align}
At larger scale factor, the scalar field will necessarily evolve. For sufficiently small ``time'' spans, we obtain the approximate solution
\begin{align}
\phi \approx \phi_0 - \sqrt{\frac{2}{3}}\mu \tau\,, \qquad a = \frac{1}{2\mu \phi_0}\, e^{\frac{\mu}{\sqrt{6}} \phi_0 \tau - \frac{1}{6}\mu^2 \tau^2}\,. \label{approxlarge}
\end{align}
where we fixed the normalisation of the scale factor by matching to the approximate solution near the origin. One may also use the approximate solution \eqref{approxlarge} to find a relation between the final values $a_1, \phi_1$ and the initial value $\phi_0,$
\begin{align}
a_1 \approx \frac{1}{2\mu \phi_0}\, e^{\frac{1}{4}(\phi_0^2-\phi_1^2)}\,. \label{aphirel}
\end{align}
This relation shows that for any specified values $a_1, \phi_1$ on the ``outer'' boundary, a suitable real initial value $\phi_0$ may be found. Moreover, the entire solution is real valued, implying that the saddle point is purely Euclidean when the potential is negative.

The action of the saddle point solutions may straightforwardly be estimated. For this, we can evaluate the on-shell action, obtained by inserting the constraint into the action,
\begin{align}
S_{on-shell}=-2\pi^2 i \int d\tau \left[ 6a + a^3 \mu^2 \phi^2\right]\,.
\end{align}
It is clear that the main contribution will come from the $a^3$ term at late times, when the universe is large. Using \eqref{approxlarge}, we immediately find 
$S_{on-shell}^{large} \approx -2\pi^2 i \sqrt{\frac{2}{3}}\mu \phi_1 a_1^3.$ But when $a$ is small, there is an additional contribution, which only becomes important after analytic continuation, but which we must include precisely for this reason. It is obtained by using the small $a$ solution \eqref{approxsmall}, and reads $S_{on-shell}^{small} \approx - 24 \pi^2 i/(\mu^2\phi_0^2).$ At the relevant saddle point, the partition function may thus be usefully approximated by
\begin{align}
\Psi(a_1,\phi_1,\phi_0) \approx e^{\frac{i}{\hbar}S_{on-shell}} \approx e^{\frac{12\pi^2}{\hbar V(\phi_0)}+\frac{2\pi^2}{\hbar}  \sqrt{\frac{2}{3}}\mu \phi_1 a_1^3}\,.
\end{align} 
In the limit where one would send the outer boundary to infinity, the divergence of the partition function with $a_1^3$ is seen to be the usual volume divergence encountered in AdS/CFT. It could be cancelled by a counter term $S_{ct}= -i \sqrt{g}\phi \partial_\tau \phi \mid_{boundary},$ see e.g. \cite{Andrade:2011dg}. 

We can now simply analytically continue $\mu$ to $-i\mu = m$ so as to find the corresponding results for the case of a positive potential. This analysis was originally done by Lyons \cite{Lyons:1992ua}, and we refer to his paper for further details. What is of prime interest to us here is the manner in which the results for a negative potential are modified by the analytic continuation. At small radii, the solution remains approximately Euclidean and is readily found, 
\begin{align}
\phi \approx \phi_0\,, \qquad a \approx \frac{\sin (\frac{m}{\sqrt{6}} \phi_0 \tau)}{\frac{m}{\sqrt{6}} \phi_0}\,. 
\end{align}
A more interesting change occurs at large radii where the solutions, which were real, now become complex; namely, we get the approximate solution
\begin{align}
\phi \approx \phi_0 - i\sqrt{\frac{2}{3}}m \tau\,, \qquad a = \frac{1}{2im \phi_0}\, e^{i\frac{m}{\sqrt{6}} \phi_0 \tau + \frac{1}{6}m^2 \tau^2}\,. \label{approxlargem}
\end{align}
We also need to consider complex values of $\phi_0,$ with the proviso that $Re(\phi_0)\gg Im(\phi_0).$ More precisely, Eq. \eqref{aphirel} implies that if we consider $Im(\phi_0)=\pi/Re(\phi_0),$ then there exists a line in the complexified $\tau$ plane where the solution becomes increasingly Lorentzian in the limit of large $a_1.$ After analytic continuation, the action becomes complex, and the wave function is given by
 \begin{align}
\Psi(a_1,\phi_1,\phi_0) \approx e^{\frac{i}{\hbar}S_{on-shell}} \approx e^{\frac{12\pi^2}{\hbar V(\phi_0)}+i \frac{2\pi^2}{\hbar}  \sqrt{\frac{2}{3}}m \phi_1 a_1^3}\,.
 \end{align}
The volume part of the action becomes real, implying that it will contribute a mere phase factor to the wave function, and no divergence any more. In fact this phase factor is welcome, as it ensures that the wave function becomes classical in a WKB sense, at large $a_1.$ The imaginary part of the action provides different weightings for different $\phi_0.$ Here we reproduce the well known result that the no-boundary wave function seems to prefer small values of the potential \cite{Hartle:2008ng,Matsui:2020tyd}, at least when one uses the naive interpretation that the probability is (approximately) given by $|\Psi|^2.$ It might be worth pointing out that a rigorous justification of this probability interpretation remains lacking, however.
 
We conclude that the equivalence between wave functions for negative and positive potentials, via analytic continuation, continues to hold for simple scalar field models. It is interesting that terms that one would eliminate on the AdS side via counter terms actually end up playing an important role once the potential is continued to positive values, as they are related to the classicality of the universe. We have only been able to check the correspondence at the level of the saddle point approximation. But it is clear that the properties of the path integral are quite similar to the case of pure gravity: once the universe grows to be large enough, for positive potentials the saddle points become complex, and thus the saddle points again appear in complex conjugate pairs. Thus a similar Stokes phenomenon as that described for a cosmological constant arises in this case also, and it is related to the emergence of time. It will be of interest to extend the current analysis to a scalar field model that can be solved exactly in minisuperspace, and where we are not limited to small excursions of the scalar field. Such a model will be presented in upcoming work \cite{upcoming}.

\section{Discussion} \label{sec:discussion}

Gravitational path integrals naturally describe transitions between two separate 3-dimensional hypersurfaces. But there are situations of interest, such as in the context of the AdS/CFT correspondence or in cosmology, where we only want to keep the metric fixed on one hypersurface. In both situations the natural boundary condition that one would like to impose at the other ``end'' of the 4-geometry is that it should round off smoothly. This is a regularity condition that turns out to be independent of the vacuum energy (or matter content) of the universe -- for the simplest boundary topology, it is the condition that the 4-geometry is locally that of a 4-sphere. As we found in Eq. \ref{noflux}, this condition also says that there is no momentum flux at the nucleation of the universe or, equivalently, that the zero size condition on the universe is imposed not in field space, but in momentum space. Thus we arrive at a succinct characterisation of the no-boundary condition,
\begin{align}
i \frac{\partial}{\partial p}\Psi = \hat{q} \, \Psi = 0 \qquad \qquad \textrm{no-boundary condition}\,,  \label{newnbc}
\end{align}
which should be satisfied in addition to the WdW equation \eqref{WdWp}. The absence of momentum flowing in or out of the universe at nucleation fits well with the no-boundary philosophy that the universe should be entirely self-contained. As a side comment, let us remark that this condition allows for a specification of the no-boundary wave function directly at the level of the WdW equation, which is in principle independent of a path integral implementation.

The no-boundary condition \eqref{newnbc} effectively removes the boundary at one end of the path integral. Not only is the condition independent of any matter content, but at the level of the action it also requires no surface term in order to allow for a consistent variational problem.  Path integrals defined in this way are essentially holographic by construction, as the only data that one is free to vary resides on the outer boundary. From this point of view, it becomes clear that the canonical partition function with asymptotically AdS boundary should be equivalent to the no-boundary wave function with asymptotically dS boundary conditions, after analytic continuation of the vacuum energy. We have verified this hypothesis for the case where tensor perturbations to the geometry are included, and also when gravity is coupled to a scalar field with a positive or a tachyonic mass term. 

It is noteworthy that when we specify the metric on one hypersurface, we have no say about the geometry in the interior spacetime. This is determined by the relevant saddle points of the path integral. In particular, we have no say even over such basic properties as to whether the resulting spacetime will be Euclidean or Lorentzian. In fact, we find that with the negative potential energies that we considered here, the saddle points always turn out to be Euclidean, i.e. we obtain 4-geometries that contain no time direction and thus no notion of causality. But when the potential is analytically continued to positive values, a Stokes phenomenon occurs in which, as the universe grows, the Euclidean saddle point splits into two complex conjugate saddles, each of which describes a geometry that at large radii becomes increasingly Lorentzian (with opposite time directions in both saddle point geometries). This is the mathematical description of how a space direction changes into a time direction. 

A corollary of the present analysis is that simple topology-changing geometries, such as a 4-sphere glued onto a no-boundary geometry, do not contribute to the gravitational path integral. Thus Hawking-Moss or Coleman-De Luccia instantons may play a role in the description of tunnelling events in a preexisting universe, but they do not seem to contribute to the creation of the universe from nothing. 

It may also be useful to contrast the present approach with the closely related, but distinct, holographic approach of Hertog and Hartle \cite{Hertog:2011ky} (for which perturbations were studied in \cite{Hertog:2015nia}). In that approach, the cosmological constant $\Lambda$ is kept fixed. However, it was noted that in the analytic continuation of the (dS) saddle point geometry, a region exists which has a geometry that is asymptotically Euclidean AdS spacetime (even though $\Lambda>0$), provided the final scale factor is large. Thus a region of the saddle point geometry effectively behaves as if $\Lambda$ had changed sign, and in this region one may apply the standard Euclidean AdS/CFT correspondence. In our notation, the total ``time'' elapsed from the nucleation of the universe until the final hypersurface is given by 
\begin{align}
T = \int_0^1 \frac{N(t)}{\sqrt{q(t)}}dt\,.
\end{align}
Whereas in the present approach $N(t)$ is taken to be constant, Hertog and Hartle split up $N(t)$ into three segments, such that one starts at the usual South Pole of the instanton and that the final time $T$ is reached, but in such a way that the intermediate segment corresponds to the Euclidean AdS geometry. The advantage of this approach is that one may make use of well established results related to the AdS/CFT conjecture and furthermore one obtains an interesting interpretation of the counter terms. What has not been demonstrated yet is whether in the path integral one may consistently split the lapse function into distinct pieces; conditions for gauge fixing the lapse function were analysed in \cite{Teitelboim:1981ua}, and the simplest way to implement these is to take $N$ constant. This point thus deserves further study. By contrast, in the approach of the present paper, the cosmological constant itself is analytically continued. What one finds is that this leads to consistent results, with AdS and dS path integrals satisfying identical (no-)boundary conditions and being related by simple analytic continuation. What remains unclear is to what extent a dual theory still exists after analytic continuation from negative $\Lambda$. We make no claims as to the existence of such a theory when $\Lambda$ is positive. Elucidating this question will require, amongst others, an understanding of AdS/CFT not just when the AdS boundary is sent off to infinity, but also at finite radius. In fact, understanding the latter issue will impact both of the approaches that have just been described.

There are a number of obvious additional questions for future investigation. The main conceptual question is perhaps the question of what the general requirements are for a Stokes phenomenon to occur, i.e. what are the circumstances necessary for the effective geometry to contain a time direction, rather than just space? Does this impose a restriction on the properties of our universe? Also, why should the canonical partition function in AdS, describing a sum over states at a fixed temperature, be at all related to the wave function of the universe? What would be the analogue concept of temperature in the latter case? Also, given the results found here, one may ask under what circumstances the correspondence between negative and positive potential energies continues to hold, or breaks down. This may be investigated in many different directions, but to give just three examples: it will be of interest to see what happens in scalar field models where the full minisuperspace version can be solved analytically, and where the scalar field is not restricted to small changes only (this is work in progress \cite{upcoming}). Furthermore, for steep negative potentials complex no-boundary instantons with late-time ekpyrotic evolution also exist \cite{Battarra:2014xoa,Battarra:2014kga}. It is not clear how these are related to corresponding solutions with positive potentials, and this may be another fruitful future research direction. Finally, in \cite{DiTucci:2020weq} an important ingredient in determining the appropriate boundary conditions for gravitational integrals with AdS asymptotics was a consideration of black holes. It would clearly be of interest to include black holes in further verifying the correspondence between AdS and dS path integrals, also with regard to the possibility of creating primordial black holes~\cite{Bousso:1996au}.

\acknowledgments

I would like to thank Alice Di Tucci for enlightening discussions that led to the formulation of the no-boundary condition in Eq. \eqref{newnbc}.
I gratefully acknowledge the support of the European Research Council in the form of the ERC Consolidator Grant CoG 772295 ``Qosmology''.

\bibliographystyle{utphys}
\bibliography{AiBi}

\end{document}